\documentclass[			
						superscriptaddress,
						twocolumn,
						a4paper,
						showkeys,
						floatfix,%
					]{revtex4-1}
\usepackage[T1]{fontenc}
\usepackage[utf8]{inputenc}
\usepackage{amsmath}
\usepackage[frenchb,english]{babel}
\usepackage{amsfonts}
\usepackage{amssymb}
\usepackage{upgreek}	
\usepackage{graphicx}
\usepackage{textcomp}	
\usepackage{hyperref} 
\usepackage{subfig}

\begin{document}
\title{Thulium and ytterbium-doped titanium oxide thin films deposited by ultrasonic spray pyrolysis}
\author{S.~Forissier}
\author{H.~Roussel}
\author{P.~Chaudouet}
\affiliation{Laboratoire des Matériaux et du Génie Physique, CNRS, Grenoble Institute of Technology, MINATEC, 3 parvis Louis Néel, 38016 Grenoble, France}
\author{A.~Pereira}
\affiliation{Laboratoire de Physico-Chimie des Matériaux Luminescents UMR 5620 CNRS / UCBL Domaine Scientifique de la Doua, Université Claude Bernard Lyon 1 Bâtiment Alfred Kastler 10 rue Ada Byron 69622 Villeurbanne cedex, France}
\author{J-L.~Deschanvres}
\email{jean-luc.deschanvres@grenoble-inp.fr}
\affiliation{Laboratoire des Matériaux et du Génie Physique, CNRS, Grenoble Institute of Technology, MINATEC, 3 parvis Louis Néel, 38016 Grenoble, France}
\author{B.~Moine}
\affiliation{Laboratoire de Physico-Chimie des Matériaux Luminescents UMR 5620 CNRS / UCBL Domaine Scientifique de la Doua, Université Claude Bernard Lyon 1 Bâtiment Alfred Kastler 10 rue Ada Byron 69622 Villeurbanne cedex, France}

\begin{abstract}
Thin films of thulium and ytterbium-doped titanium oxide were grown by metal-organic spray pyrolysis deposition from titanium(IV)oxide bis(acetylacetonate), thulium(III) tris(2,2,6,6-tetramethyl-3,5-heptanedionate) and ytterbium(III) tris(acetylacetonate). Deposition temperatures have been investigated from 300\textdegree C to 600\textdegree C. Films have been studied regarding their crystallity and doping quality. Structural and composition characterisations of TiO$_{2}$:Tm,Yb were performed by electron microprobe, X-ray diffraction and Fourier transform infrared spectroscopy. The deposition rate can reach 0.8 $\upmu$m/h. The anatase phase of TiO$_{2}$ was obtained after synthesis at 400\textdegree C or higher. Organic contamination at low deposition temperature is eliminated by annealing treatments.

\end{abstract}

\maketitle

\section*{Introduction}
Over the last decades, titanium dioxide has been attracting great interest due to their relevance for a variety of applications. For example, TiO$_{2}$ has been investigated for use in photocatalytic process\cite{Diebold2003,Carp2004,Foster2011}, in protective coatings\cite{Poulios1999} and gas sensing applications\cite{Savage2001}. TiO2 is also interesting for photovoltaic applications\cite{Richards2004}. TiO2 thin films have been the most commonly used antireflective (AR) coating in the photovoltaic industry. Owing to its excellent optical properties, mechanical properties and good chemical resistance, TiO2 still remains attractive for such application.

The physical and chemical properties of oxide matrix can be tuned by doping with rare-earth (RE) elements. For example, the incorporation of RE ions into oxide matrix received great attention for applications in photovoltaic devices\cite{Zhang2010,Ende2009,Guille2012}. Indeed, one solution to reduce the energy losses in the UV region due to thermalization of the charge carriers is to adapt the solar spectrum to better match the bandgap of solar cells. This approach, which involves energy transfer between rare-earth ions, is well known for lighting issues\cite{Zhang2010,Beauzamy2008} and has been recently proposed for photovoltaic applications\cite{Richards2006a}. Depending on both the host matrix and the RE ions used, down-shifting or quantum-cutting can be observed. In this context, RE-doped titanium dioxide can be a potential low-cost material for the down-conversion. The use of titanium oxide being well-established in the photovoltaics industry, added functionality with rare-earth could be easily integrated. However, depending on the size of the RE ions such doping can be challenging. It is thus expected that the choice of the RE ions, the method used to synthesize the RE-doped TiO2 and the experimental parameters prevailing during the synthesis will have a strong influence on the final material.

Most studies concerning rare-earth doped titanium oxide have been performed on thin films, which have been fabricated by sol-gel\cite{Frindell2003,Hu2007,Saif2007,Jia2006,Amlouk2008} and sputtering processes\cite{Prociow2007}. In most of these case, the studies were mainly concerned with samarium\cite{Frindell2003,Hu2007,Saif2007}, europium\cite{Frindell2003,Saif2007,Prociow2007}, terbium\cite{Frindell2003,Saif2007,Jia2006}, erbium\cite{Frindell2003}, neodymium\cite{Frindell2003},  praseodymium\cite{Amlouk2008} and ytterbium\cite{Frindell2003}. All of these can be used as photocatalysts or optical layers for light applications. Up to now, few reports have appeared in the literature concerning spray-pyrolysis synthesis of RE-doped titanium dioxide thin films\cite{Kanarjov2008}. This technique is well known as a powerful method for the synthesis of all kind of materials (e.g. fluorides, oxides, metals) at atmospheric pressure. Moreover, it allows excellent control of the elemental composition of the resultant films through the addition of dopant elements in the desired ratio to the solution medium. Thin films with large area can thus be obtained with a simple technique at relatively low cost and easily scalable (e.g. roll to roll processes). Previous studies have demonstrated the possibility of titanium oxide synthesis by this technology\cite{Castaneda2003,Conde-Gallardo2005,Duminica2007,Senthilnathan2010}. For example, Castaneda et al. and Conde-Gallardo et al. have shown by X-ray diffraction that crystallization in anatase phase occurs over 300-400\textdegree C, \cite{Castaneda2003,Conde-Gallardo2005} whereas Duminica et al. have reported a fraction of rutile phase at high temperature \cite{Duminica2007}.

In this study, we investigated the structural and spectroscopic properties of thulium and ytterbium-co-doped titanium dioxide films prepared by spray pyrolysis.  The influence of different deposition parameters on the doping level and the crystalline structure of the films were investigated. Energy transfer mechanisms in the co-doped TiO2 films are also discussed. 

\section{Experimental}
Thulium and ytterbium-doped titanium dioxide thin films were deposited by mean of ultrasonic spray pyrolysis method\cite{Mooney1982}. The precursor solution, was composed of titanium(IV) oxide bis(acetylacetonate), TiO(acac)$_{2}$, thulium(III) tris(2,2,6,6-tetramethyl-3,5-heptanedionate) and ytterbium(III)(acac) dissolved in high purity butanol (99\%) at a total concentration of 0.03 mol/l. All these precursors were purchased from STREM Chemicals Inc (Bischheim, France) and butanol was purchased from Alfa Aesar GmbH (Schiltigheim, France). Precursors were selected for non-toxicity, good stability at room temperature, easy handling, high volatility and low cost \cite{Ryabova1968}. The films were deposited on (100) silicon substrate. A similar ultrasonic spray pyrolysis apparatus used in this study is presented in \cite{Deschanvres1990,Deschanvres1993}. However this study's apparatus has a vertical sampler holder instead of a horizontal one.

The aerosol was produced by means of a flat piezoelectric transducer excited at 800~kHz which generated an ultrasonic beam in a solution containing the reactant of the material to be deposited. The source solution is delivered to the piezoelectric transducer through a constant level burette to ensure a constant rate of vapour formation during deposition. This ultrasonic spraying system guarantees a narrow dispersion of the droplet size, and based on the experimental formula by Lang\cite{Lang1962} the size of the droplets is 3.6~$\upmu$m.

Droplets are carried with two air fluxes, dried and purified, to the heated sample holder, the lower air flux (12.7~l/m) is the main carrier gas while the upper air flux (10.1~l/m) lengthens the time that the vapour stay in the vicinity of the sample holder. The overall air flow is parallel to the substrate's surface. For deposition, the substrate was fixed by clips on the sample holder heated by an electrical resistance with a built-in thermocouple. Due to this narrow distribution, in an appropriate temperature range the pyrolysis of the aerosol correspond to CVD process as described by Spitz and Viguié\cite{Viguie1975}.

The X-ray diffraction profiles were obtained with a Bruker D8 Advance using Cu $K\alpha_1$ radiation at 0.15406~nm with an applied voltage of 40~kV and 40~mA anode current in $\theta/2\theta$ configuration. Absorption FT-IR spectroscopy was used to study the structural evolution of the films versus the deposition conditions. Spectra were obtained between 250 and 4000~cm$^{-1}$ with 4~cm$^{-1}$ resolution with a Bio-Rad Infrared Fourier Transform spectrometer FTS165 and after performing Si substrate subtraction. SEM pictures were taken with a FEI Quanta FEG 250 microscope using an Everhart-Thornley detector in the secondary electron mode. The AFM pictures were taken with a Veeco Dimension 3100 with a Quadrex equipment. The composition of the doped films were measured by electron probe microanalysis (EPMA Cameca SX50) and computed by help of special software dedicated to the thin film analysis, called Stratagem and edited by the SAMx society \cite{Pouchou1984}. Luminescence spectra were taken using a F900 spectrofluorimeter Edinburgh with a high spectral resolution. A Xenon Arc lamp (450 W) is used for the excitation, the detector is a photomultiplier Hamamatsu R2658P cooled by Peltier effect.

\section{Results and discussion}
In a first step we compared the deposition temperatures' effect on the samples' composition and morphology; then their structural properties are analysed. The last step will expose some of the spectroscopic properties of the thin films.

\subsection{Morphology and composition}

\begin{figure}[!ht]
	\includegraphics[width=0.47\textwidth]{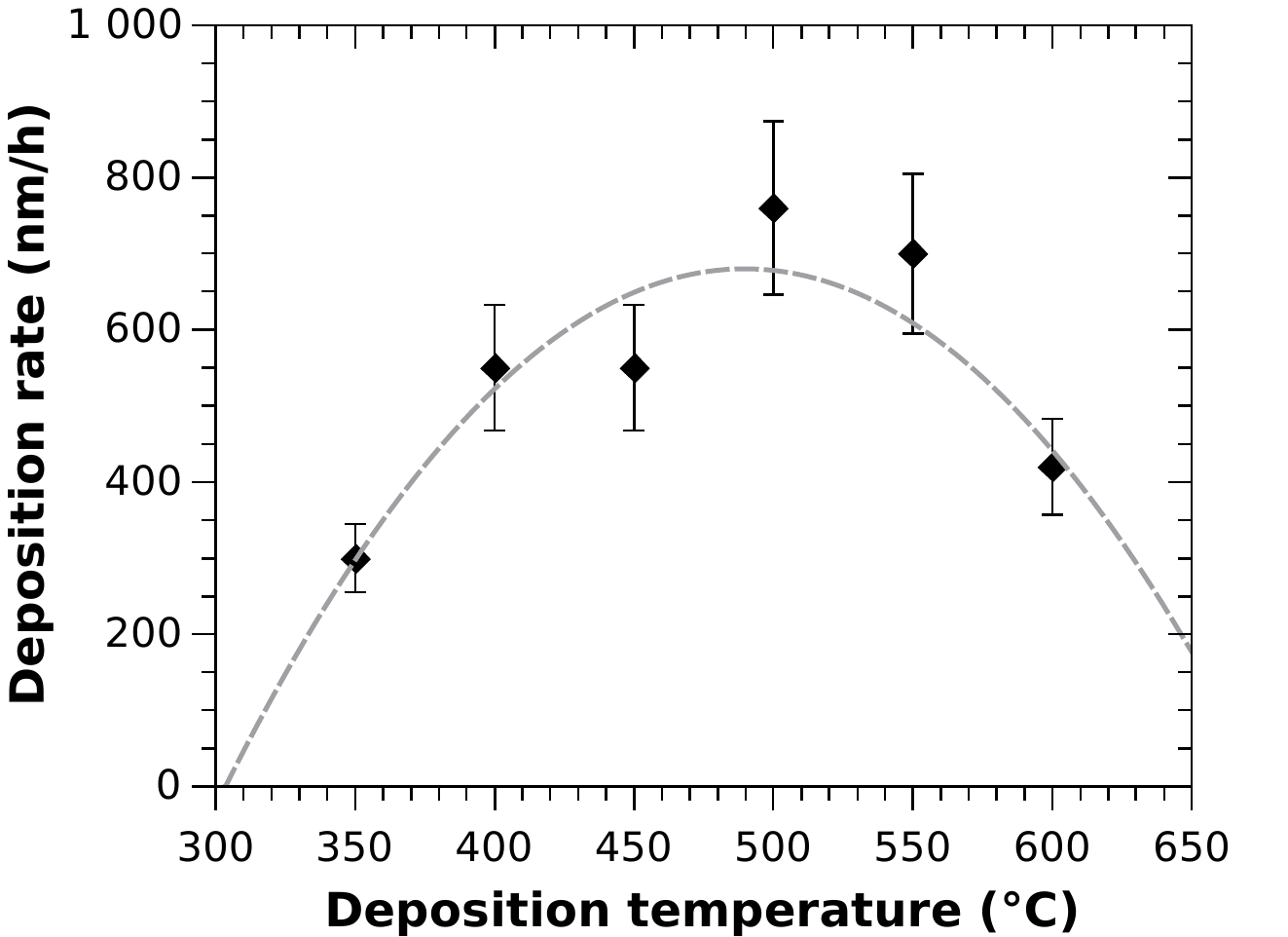}
	\centering
	\caption{Deposition rate versus temperature.}
	\label{vitdep}
\end{figure}

First we studied the deposition rate of the thin film measuring their thickness by SEM cross-section. We observe a increase of this rate until 500\textdegree C with a maximum value of 760 nm/h (Figure\ref{vitdep}). Above 500\textdegree C the decrease of the deposition rate correspond to the depletion regime where the precursor vapour reacts before reaching the substrate and produced non-adherent thin films.

\begin{figure}[!ht]
	\centering
	\subfloat[\label{comptemp}]{\includegraphics[width=0.47\textwidth]{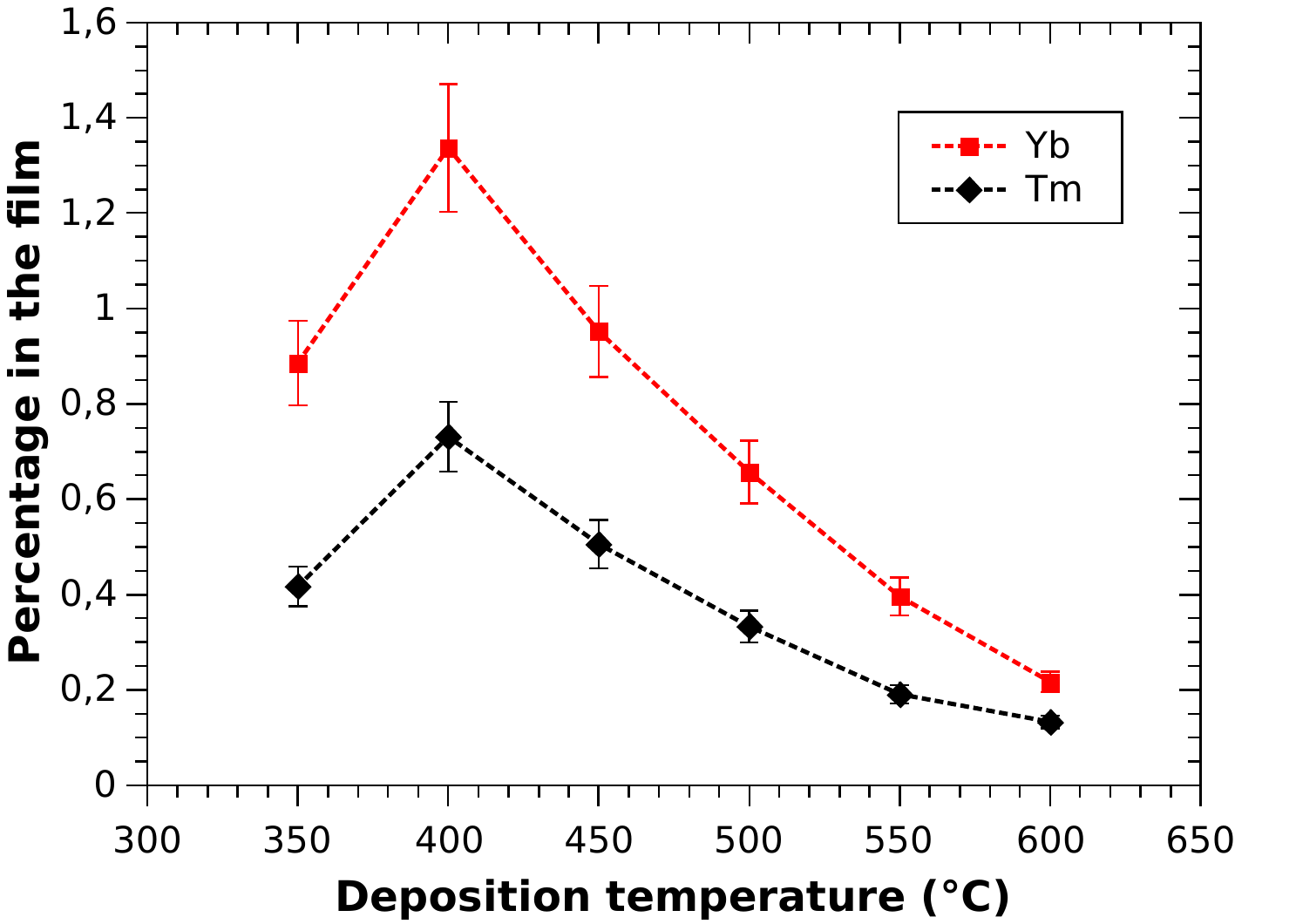}}\\
	\subfloat[\label{compgene}]{\includegraphics[width=0.47\textwidth]{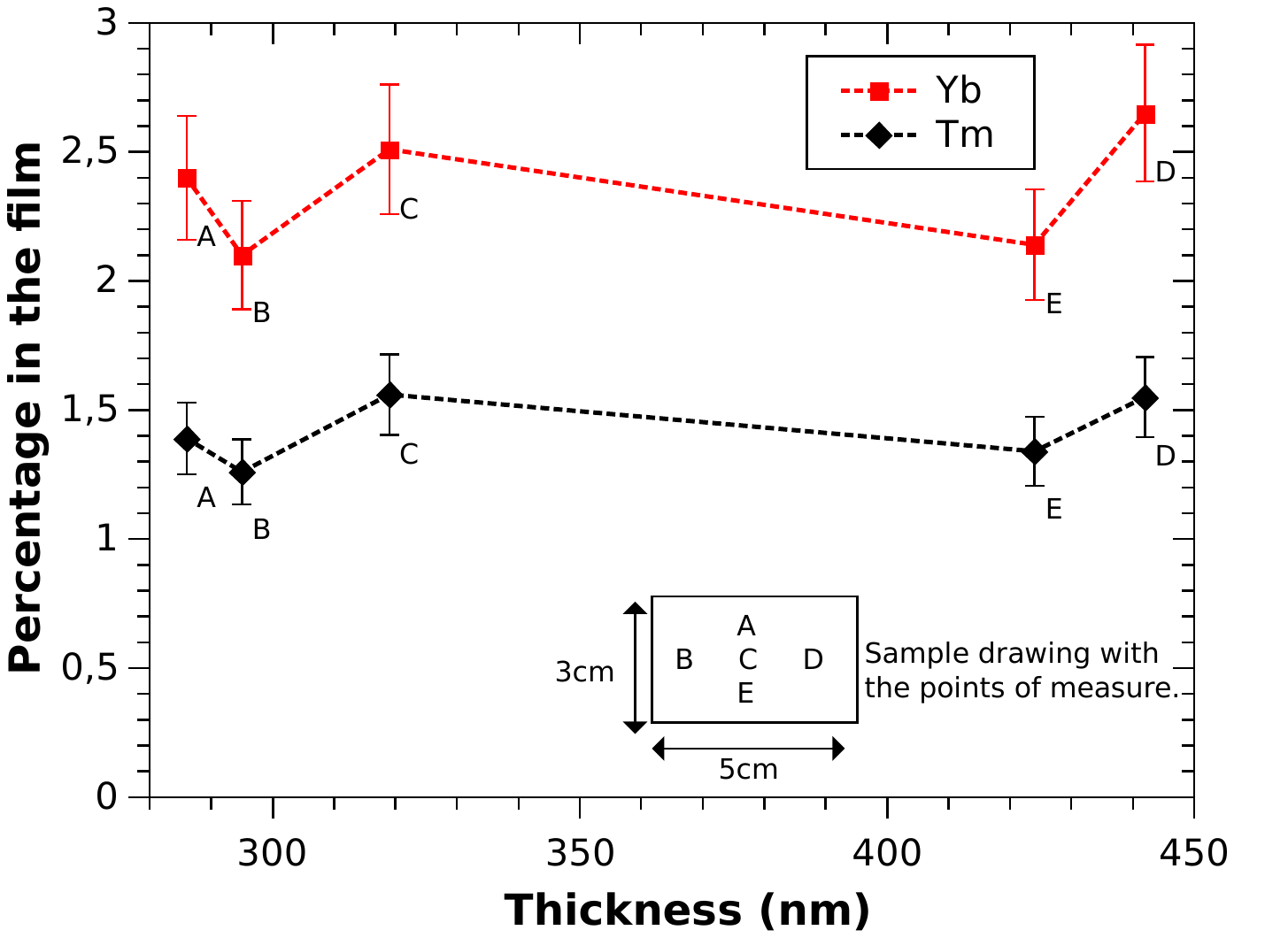}}
	\caption{\ref{comptemp}) Rare-earth cationic concentration ($\frac{RE}{\Sigma RE+Ti}$) in the film versus deposition temperature from an original solution which contained a  concentration of 3\% of Tm and 3\% of Yb. \hspace*{8ex}\ref{compgene}) Rare-earth cationic concentration and thickness (measured by EPMA and analysed by Strata) of the films deposited at 400\textdegree C at different points on the substrate surface (drawing). Initial solution : 5\% Tm and 6\% Yb.}
\end{figure}

Then we synthesized different samples at different temperature from the same solution with a rare-earth cationic concentration of 3\% of Tm and 3\% Yb in the solution ($\frac{RE}{\Sigma RE+Ti}*100$); rare-earth concentration in the film measured by electron microprobe, measurement of this batch are shown on Figure~\ref{comptemp}. As we noticed that the rare-earth concentration in the films is always lower than 3\%, the rare-earth precursor reactivity is shown to be less than that of the titanium oxide precursor. The Yb precursor reactivity is higher that the one of the Tm precursor and the highest doping efficiency from the solution to the film is obtained at 400\textdegree C. 

Subsequent experiments were mainly conducted at this temperature and even in this condition the rare-earth precursor reactivity is lower than the titanium oxide precursor reactivity. Due to the reactor's geometry, the thickness of the sample are not homogeneous on the whole surface. Nevertheless the doping level uniformity is quite good on the whole surface as shown on Figure~\ref{compgene} (drawing).

\begin{figure}[!ht]
	\centering
	\qquad
	\subfloat[AFM surface view.\label{afm}]{\includegraphics[width=0.47\textwidth]{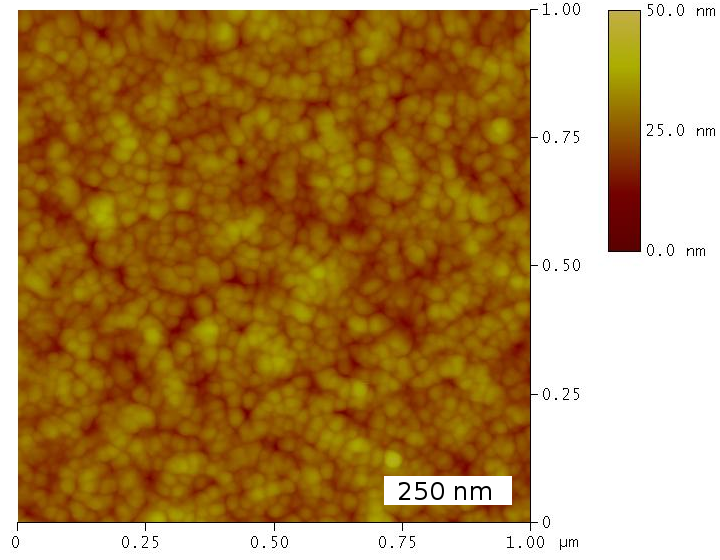}}\\
	\subfloat[SEM cross-section view.\label{feg}]{\includegraphics[width=0.47\textwidth]{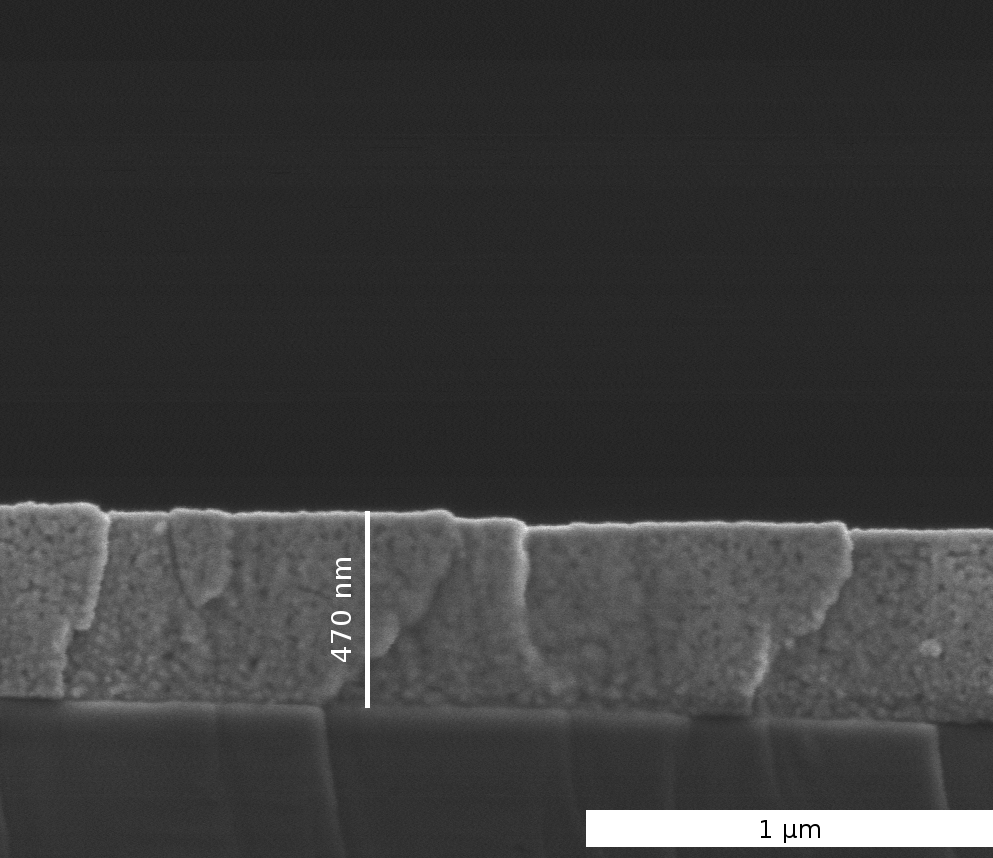}}
	\qquad
	\caption{Morphological pictures of a annealed sample at 800\textdegree C.}
	\label{morph}
\end{figure}

The as-deposited and even annealed sample exhibit a smooth surface as shown by AFM picture (Figure~\ref{afm}) with a measured RMS roughness of less than 4 nm for a film thickness of 420 nm (measured by EPMA). The SEM cross-section picture shows a good density in the thin films and a thickness of the same order.

\begin{figure}[!ht]
	\includegraphics[width=0.47\textwidth]{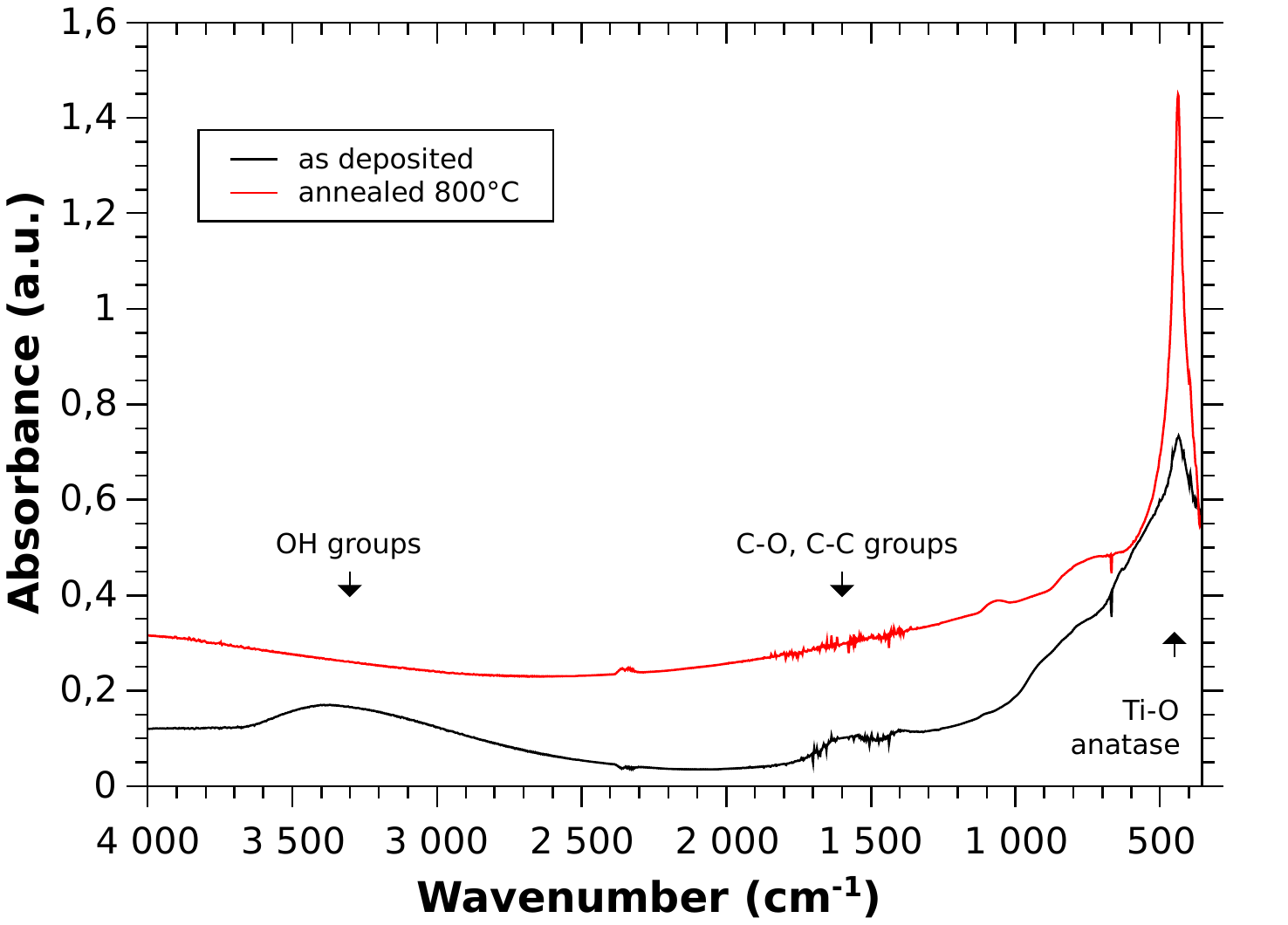}
	\centering
	\caption{Infrared spectra of as-deposited and annealed sample. Organic residues are disappearing with the annealing.}
	\label{ftir}
\end{figure}

FT-IR measurements (Figure~\ref{ftir}) showed that as-deposited thin films still have some organic residues with typical hydroxyl band (3000--3500 cm$^{-1}$) along with C-O and C-C bonds signatures (1700--1300 cm$^{-1}$). Those organic residues of the precursors are eliminated by post-annealing treatment (800\textdegree C for 1h).

\subsection{Structural properties}

\begin{figure}[!ht]
	\centering
	\subfloat[As-deposited samples at different synthesis temperatures.	\label{DRX_deposition}]{\includegraphics[width=0.45\textwidth]{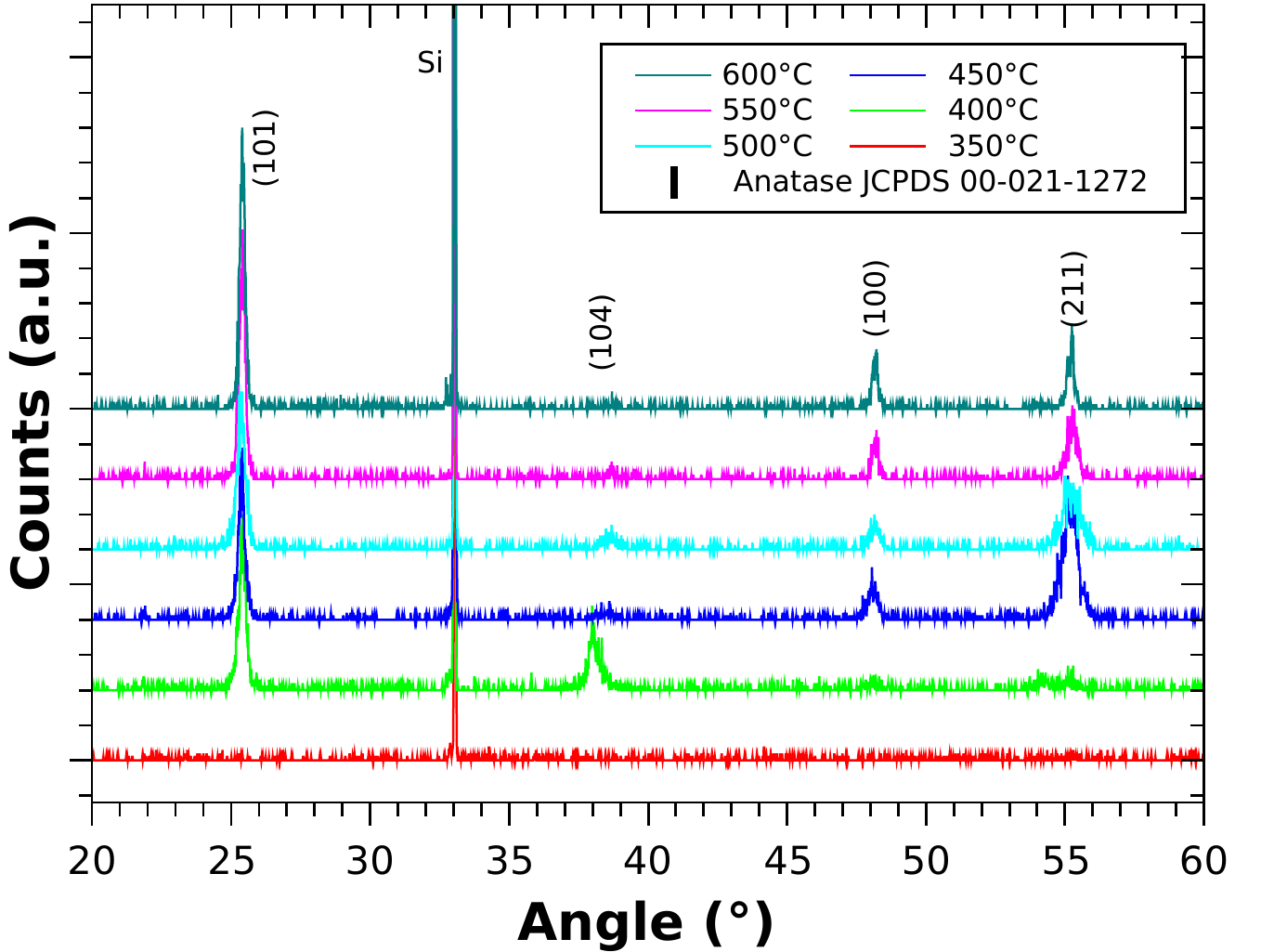}}\\
	\subfloat[As-deposited and annealed samples for 1h at 500\textdegree C or 800\textdegree C.\label{DRX_annealing}]{\includegraphics[width=0.45\textwidth]{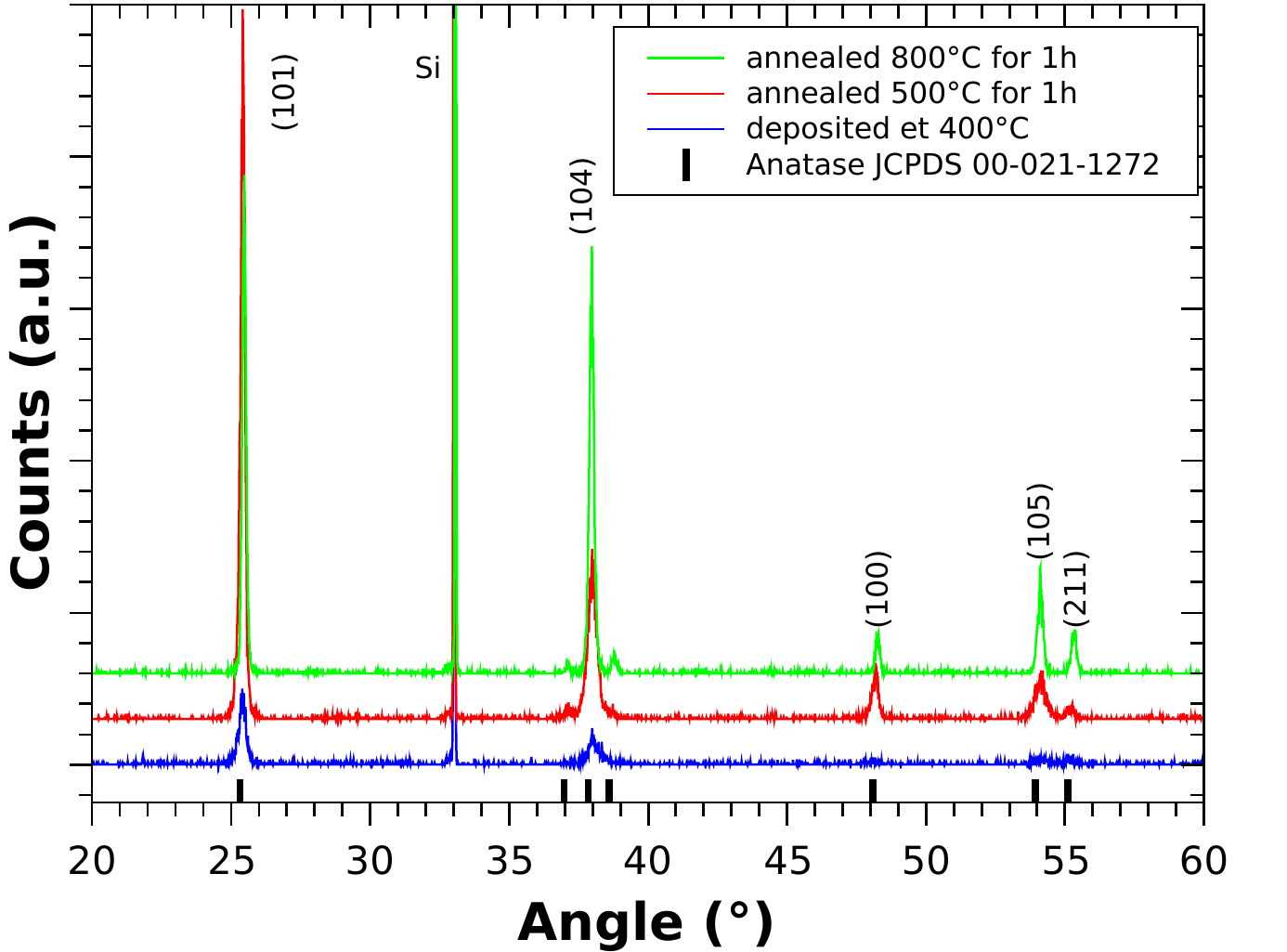}}
	\caption{X-ray diffraction spectra.}
\end{figure}

XRD spectroscopy showed that samples are amorphous when deposited below 400\textdegree C (Figure~\ref{DRX_deposition}). From 400\textdegree C to 600\textdegree C,  we observe the anatase phase of titanium oxide even for total rare-earth concentration up to 8\%. On the basis of the literature, we assume that the RE ions enter the anatase structure as substitutional defects with respect to Ti (\cite{Ghigna2007}). We tried different annealing conditions, first 500\textdegree C for 1h then 800\textdegree C for 1h in air; each annealing improves the crystalline quality of the films as shown on Figure~\ref{DRX_annealing}. With increasing annealing temperature the anatase peaks exhibit higher intensity and smaller width. After annealing at 800\textdegree C RE-doped samples are still crystallised in the anatase phase contrary to undoped samples which are crystallized in the rutile phase. The presence of rare-earth dopants prevents the phase transition as reported in \cite{Graf2007}.

\subsection{Luminescence properties}
\begin{figure}[!ht]
	\centering
	\subfloat[Emission spectra, both ions are luminescing.
	\label{lum-em}]{\includegraphics[width=0.47\textwidth]{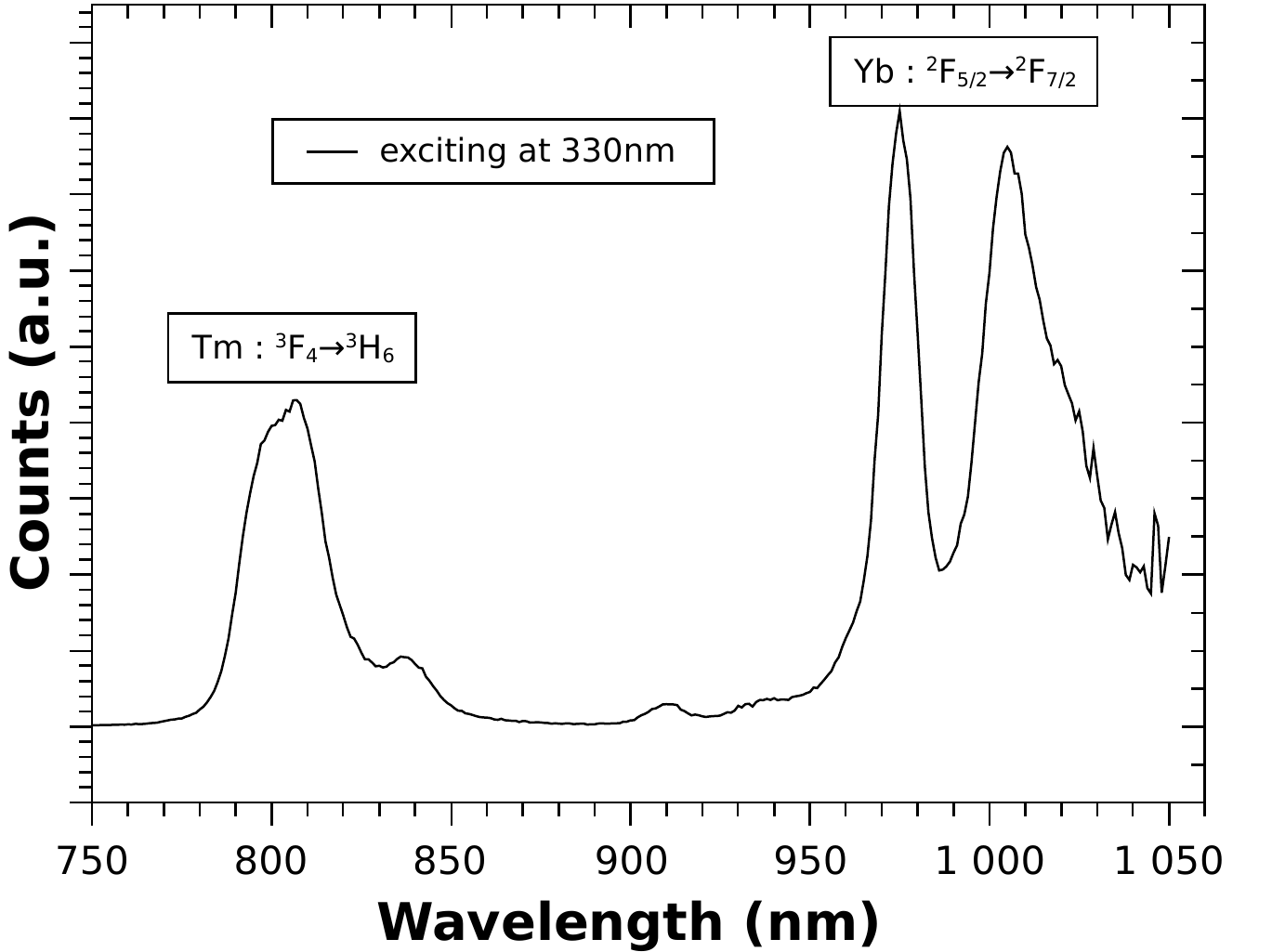}}\\
	\subfloat[Excitation spectra, the thin film is absorbing light through the matrix.
	\label{lum-ex}]{\includegraphics[width=0.47\textwidth]{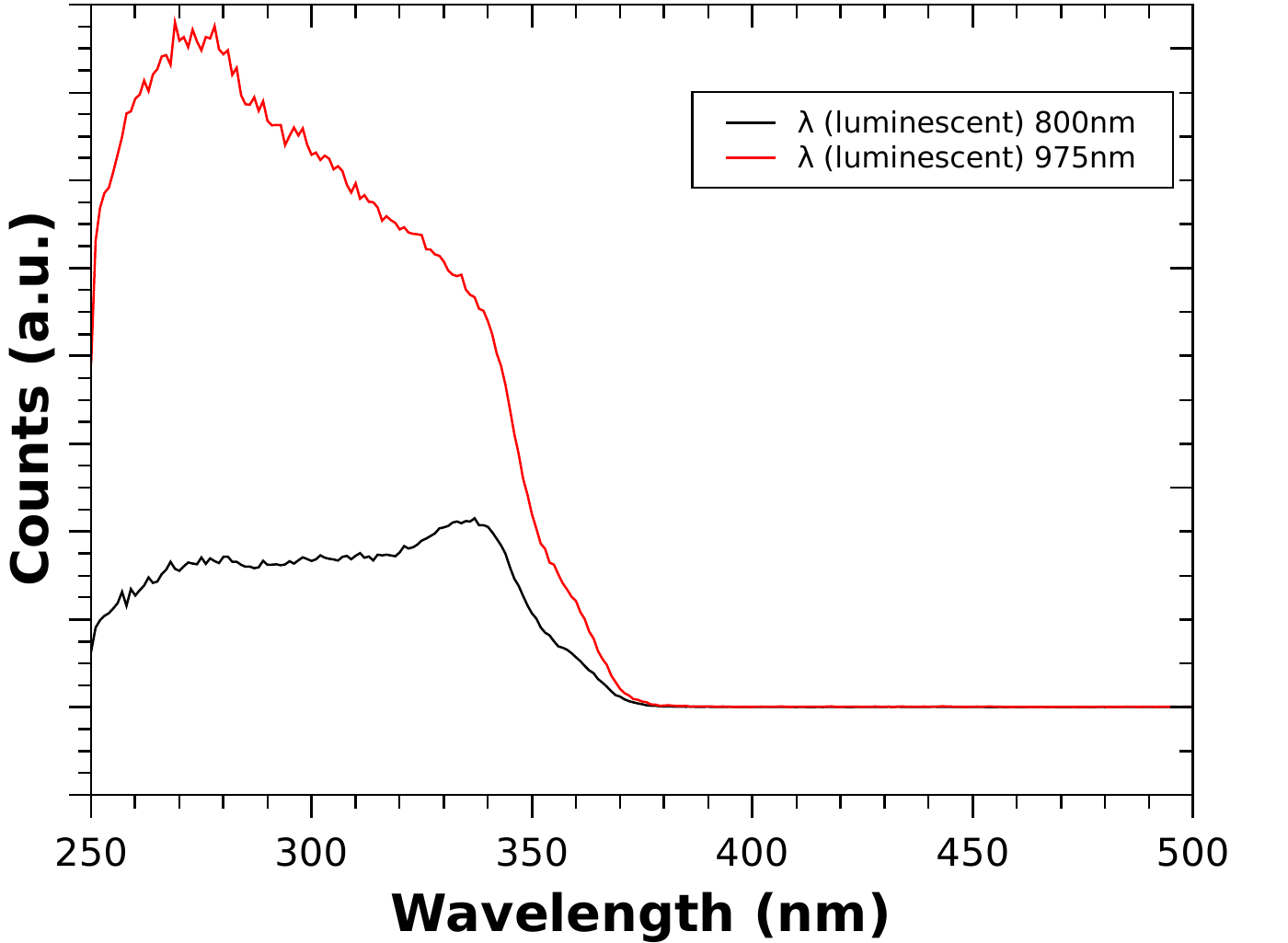}}
	\caption{Luminescence spectra of a Tm and Yb-doped TiO$_2$, respectively 0,37\% and 0,85\% measured by EPMA.}
\end{figure}

Several RE couples have been studied for down-conversion, most of them are doped with ytterbium as emitter and with another RE as absorber such as praseodymium\cite{Chen2008}, terbium\cite{Ting2003,Das2011} or europium\cite{Strek2000}. The thulium and ytterbium couple luminescence in oxide matrix is also known in the literature, Dominiak-Dzik et al have shown energy transfer between the RE ions\cite{Dominiak-Dzik2008}.

The emission scans (Figure~\ref{lum-em}) were recorded exciting at 330~nm. At this wavelength the excitation is near the $^1D_2$ level of thulium and absorbed by the TiO$_2$ matrix. On mono-doped samples (not figured) we see the respective ion transition : the Tm $^3F_4-^3H_6$ luminescence around 800~nm and the Yb $^2F_{5/2}-^2F_{7/2}$ luminescence at 980~nm. On Tm,Yb co-doped samples (Figure~\ref{lum-em}) both transitions are recorded. The exciting energy appears to be transferred to both RE ions through the TiO$_2$ matrix leading to a down-conversion mechanism with thulium and ytterbium. The Tm luminescence is well inside the absorption range and the Yb luminescence is right before the band gap of the silicon solar cells.

We recorded excitation scans for both above-mentioned Tm and Yb transitions (Figure~\ref{lum-ex}). Their luminescence is triggered through near-UV absorption between 300 and 350~nm. This corroborates the idea of the energy transfer between the matrix and the rare-earth ions.

\section*{Conclusion}
In summary, we succeeded in depositing smooth and uniform film of doped titanium dioxide with thulium and ytterbium (up to 8\%) by the ultrasonic spray pyrolysis method and to grow them partially crystallized mainly in the anatase phase. The crystallization is improved after proper annealing.

\section*{Acknowledgement}
The authors would like to thank Mr. Olivier Raccurt from the CEA-Liten Grenoble for providing access to the spectrophotometer.

Funding for this project was provided by a grant from the French Research National Agency (ANR) through Habitat intelligent et solaire photovoltaïque program (project MULTIPHOT n° ANR-09-HABISOL-009), the CARNOT institute Energie du futur and la Région Rhône-Alpes. Sébastien Forissier held a doctoral fellowship from la Région Rhône-Alpes.

\bibliographystyle{unsrt}
\bibliography{/home/forissis/these/biblio/biblio,/home/forissis/these/biblio/bibgermain}

\end{document}